\documentstyle[12pt]{article}
\begin{document}
\begin{titlepage}
\hfill   SLAC-PUB-6495

\hfill   May 1994

\hfill   T/E
\vfill
\begin{center}
\bf\large
Light-Ray Operators and their Application in QCD
\end{center}
\vspace{0.25in}
\begin{center}
B. Geyer, D. Robaschik\\
\vspace{0.2in}
Fachbereich Physik der Universit\"at Leipzig, Germany\\
\vspace{0.2in}
D. M\"uller\footnote{Supported in part by
                     Deutschen Akademischen Austauschdienst
              and by Department of Energy contract DE-AC03-76SF00515.}\\
\vspace{0.2in}
 Stanford Linear Accelerator Center,\\
 Stanford University, California 94309
\end{center}
\vspace{0.25in}
\begin{center}
{\it Presented at the Leipzig Workshop on }\\
{\it Quantum Field Theoretical Aspects of High Energy Physics,}\\
{\it Bad Frankenhausen, Germany, September 20-24, 1993.}
\end{center}
\vfill
\begin{abstract}
The nonperturbative  parton  distribution  and  wave  functions  are
directly related to matrix elements of light-ray (nonlocal) operators.
These  operators  are  generalizations of the standard local operators
known  from the operator product expansion. The renormalization group
equation  for  these  operators  leads  to evolution equations for more
general distribution amplitudes which include the Altarelli-Parisi and
the  Brodsky-Lepage  equations  as  special  cases. It is possible to
derive  the Altarelli-Parisi kernel as a limiting case of the extended
Brodsky-Lepage   kernel.   As  new  application of the  operator product
expansion the virtual Compton scattering near forward direction is
considered.
\end{abstract}
\vspace{0.25in}
\begin{center}
\mbox{\ }
\end{center}
\vfill
\end{titlepage}
\thispagestyle{empty}
\newpage

\section{Light-Ray Operators in QCD}
\setcounter{equation}{0}

It is generally accepted that nonlocal composite operators are
necessary to describe hadrons in QCD. For example a meson
operator $ M(x_1,x_2) $ could be represented by quark-antiquark
fields connected by a path-ordered phase factor $ U(x_2, x_1) $ :
\begin{eqnarray}
\label{v.1}
   M(x_1, x_2) = \overline\psi(x_2) \Gamma U(x_2, x_1) \psi(x_1)
\end{eqnarray}
with
\begin{eqnarray}
\label{v.2}
      U(x_2, x_1) = P exp(-ig \int_{x_1}^{x_2} dx^\mu A_\mu(x));
\end{eqnarray}
where $\Gamma$ abbreviates $\gamma$- and $\lambda$-matrices describing
the spin and flavor structure of the meson. $A^{\mu }=A^{\mu }_a t^a$
denotes the gluon field, and the $t^a$ are the generators of the colour
group.
For  large  internal  momenta  the  main contribution comes from those
operators for which the quark fields and the path-ordered phase factor
lie on  a light-like straight line $(x_1 -x_2)^2 = 0 $. Such operators
are known  from the Operator Product Expansions \cite{OP}.
They are highly singular  and  posses  new  properties.
For simplicity let us write the corresponding  Light-Cone Expansion for
a toy model with scalar fields $\phi$ and scalar currents $j$:
\begin{eqnarray}
\label{v.3}
 Tj(x)j(0) =  \sum_{n_1, n_2} C_{n_1, n_2} O_{n_1,n_2} + ....
             \hspace{0.5cm},
\end{eqnarray}
where the $C_{n_1,  n_2}(x^2)$  are singular coefficient functions, and
$O_{n_1,n_2 }$ are the operators
\begin{eqnarray}
\label{v.4}
   O_{n_1,n_2 } = :[(\tilde x \partial_x)^{n_1} \phi(x)]
                   [(\tilde x \partial_x)^{n_2} \phi(x)] |_{x=0}:
                   \hspace{0.5cm},
\end{eqnarray}
and $\tilde x$ is a light-like vector that approaches the vector $x$.
This  local  light-cone  expansion  can  be  summed  up  to a nonlocal
integral representation
\begin{eqnarray}
\label{v.5}
 Tj(x)j(0) = \int_{0}^{1}d\kappa_1 d\kappa_2 F(x^2,\kappa_1, \kappa_2)
             O(\kappa_1,\kappa_2 ) + ....  ,
\end{eqnarray}
where
\begin{eqnarray}
\label{v.6}
   O(\kappa_1, \kappa_2) & = & : \phi(\kappa_1\tilde x)
                                 \phi(\kappa_2\tilde x): \\
\label{v.7}
                  & = & \sum_{n_1, n_2}\frac{\kappa_1^{n_1}}{n_1!}
                                       \frac{\kappa_2^{n_1}}{n_2!}
                         O_{n_1,n_2 }
\end{eqnarray}
and
\begin{eqnarray}
\label{v.8}
              C_{n_1,  n_2}(x^2) ={1\over n_1! n_2!}
            \int_{0}^{1}d\kappa_1 d\kappa_2 F(x^2,\kappa_1, \kappa_2)
             {\kappa_1^{n_1}}
             {\kappa_2^{n_1}}.
\end{eqnarray}
Such a  nonlocal   light-ray   operator  product  expansion  has  been
independently introduced by \cite{AZ} and \cite{BB} from different points of
view.
The  operators  $ O(\kappa_1, \kappa_2) $ reflect the renormalization
properties of all summed up local operators.
The  support  of this operator is a light-like straight line
which  introduces  new singularities and is responsible for the changed
renormalization properties.
The  operator  $  O(\kappa_1,\kappa_2)$  depends on two parameters
$\kappa_1$ and $\kappa_2$ so that in a multiplicative
renormalization  scheme  the  Z-factors  and  the anomalous dimensions
depend on two initial  and two final parameters, i.e.,
\begin{eqnarray}
\label{v.9}
      O(\kappa_1, \kappa_2)^{ren} =
          \int_{0}^{1}d{\kappa'_1} d{\kappa'_2}
           Z(\kappa_1,  \kappa_2 ;  {\kappa'_1},  {\kappa'_2} )
      O({\kappa'_1}, {\kappa'_2})^{un},
\end{eqnarray}
\begin{eqnarray}
\label{v.10}
\lefteqn{\gamma (\kappa_1, \kappa_2;  {\kappa'_1},{\kappa'_2} ) =}
\\
  & & {1\over 2}\int_{0}^{1}d{\kappa''_1} d{\kappa''_2}
\left(\mu {d \over d\mu}Z (\kappa_1, \kappa_2; {\kappa''_1}, {\kappa''_2})
\right)
    Z^{-1}({\kappa''_1},  {\kappa''_2}; {\kappa'_1}, {\kappa'_2}),
\nonumber
\end{eqnarray}
where $\mu$ denotes the renormalization point.
It is possible to derive the standard local anomalous dimensions from
the nonlocal anomalous dimension.\\
In a very formal way we may write the renormalization group equation for
the operator $ O(\kappa_1, \kappa_2)^{ren}(\mu) =
RTO (\kappa_1,\kappa_2) exp\{iS_I\} $ according to
\begin{eqnarray}
\label{3.2}
\lefteqn{ \mu {d \over d\mu}
     RTO (\underline{\kappa }) e^{iS_I} = } \\
   & &   \int\!d^2
\underline{\kappa'} \left( \gamma (\underline{\kappa },
\underline{\kappa' };g(\mu ) ) - 2\gamma_\psi( g(\mu )) \delta^{(2)}
(\underline{\kappa }- \underline{\kappa' }) \right)
RTO(\underline{\kappa' }) e^{iS_I} , \nonumber
\end{eqnarray}
where $R$ denotes the renormalization procedure, $T$ the
time ordering, and $S_I$ is the interacting part of the action.
The new variables are $\kappa_\pm =(\kappa_2 \pm \kappa_1)/2$.
For convenience we split the anomalous dimension of the operator
into the anomalous dimension of the 1PI vertex function
(with two external momenta)
$\gamma (\underline{\kappa } ,\underline{\kappa ' }) =  \gamma
(\kappa_+,\kappa_-,\kappa '_+ ,\kappa '_- ) $
and a part which is proportional to the anomalous dimension of
the quark field $ \gamma_\psi$. As short notations we use
$d^2 \underline{\kappa '}=d\kappa'_+ d\kappa '_-$, and $\delta^{(2)}
(\underline{\kappa  }-\underline{\kappa '  }) = \delta (\kappa _+ -
\kappa '_+)\linebreak[2] \delta (\kappa_- - \kappa '_-)$.

\section{ Generalized Distribution Functions and Evolution Equations}
\setcounter{equation}{0}
\renewcommand{\theequation}{\arabic{section}.\arabic{equation}}

Before  defining  generalized  distribution  functions  we  repeat the
definitions  of  the  well-known  meson  wave  function and the parton
distribution function in QCD.

The meson wave function $\Phi^a (x,Q^2)$ which depends on
the distribution parameter $x$ $(0 \leq x \leq 1)$  can be defined
as the expectation value of a nonlocal (light-ray) operator lying on the
light-cone \cite{BL,ZC,Ge}:
\begin{eqnarray}
\label{2.1}
\lefteqn{\Phi^a (x=\frac{1+t}{2},Q^2)=}\\
& &\int\!{d\kappa_- (\tilde n P) \over 2\pi
(\tilde nP)}\;e^{i\kappa_-(\tilde nP)t} <0|RTO^a (\kappa_-;\tilde n)
e^{iS_I}|P>|\scriptstyle {\mu}^2=Q^2 \displaystyle.
\nonumber
\end{eqnarray}
Here $|P>$ denotes the one-particle state of a scalar meson of momentum $P$,
and
\begin{eqnarray}
\label{2.2}
O^a (\kappa_- ;\tilde n) =\; :\!\bar \psi(-\kappa_- \tilde  n)  (\tilde  n
\gamma) \lambda^a U(-\kappa_- \tilde n,\kappa_- \tilde n) \psi(\kappa_-
\tilde n)\!:
\end{eqnarray}
is the light-ray operator with the same flavor content.
As the renormalization point we choose the typical large momentum $Q$ of
the basic process to which the wave function contributes.
The introduced factor $1/(\tilde n P)$ compensates the $\tilde n$-dependence
of the factor $\tilde n \gamma$ in (\ref{2.2}).
The vector $\tilde n$ with
${\tilde n}^2 = 0$ defines the light-ray pointing in the direction of the
large  momentum flow of the process. The path ordered phase factor is
taken along the straight line with direction $ \tilde n$, and
$\lambda^a$ is a generator of the flavor group
corresponding to the considered meson.

Analogously we define the quark distribution function $q^a(z,Q^2)$ \cite{AP}
with the distribution parameter $z$.
For simplicity we consider the flavor nonsinglet distribution only defined by
\begin{eqnarray}
\label{2.5}
q^a (z,Q^2)= \int\!{d\kappa_- (\tilde n P) \over 2\pi (\tilde n P)}
\;e^{2i\kappa_- (\tilde n P)z}<P|RTO^a (\kappa_- ;\tilde n)
e^{iS_I}|P>|\scriptstyle {\mu}^2=Q^2 \displaystyle.
\end{eqnarray}
If we choose the index $a$ so that the matrix $\lambda^a$ is diagonal,
then  for a positive resp. a negative distribution parameter $z$ this
function  represents a linear combination of the quark resp. antiquark
distribution functions \cite{AP}.

The functions in (\ref{2.1}) and (\ref{2.5}) are used for different physical
processes and have different interpretations. Nevertheless it seems to be
natural to introduce the more general distribution amplitude
\begin{eqnarray}
\label{2.6}
\lefteqn{q^a (t,\tau ,\mu^2)=}\\
& &\int\!{d\kappa_- (\tilde n  P_+)\over 2\pi
(\tilde nP_+)} \;e^{i\kappa_- (\tilde n P_+)t}<P_2|RTO^a (\kappa_- ;\tilde
n) e^{iS_I}|P_1>|\scriptstyle {\tilde n P}_-=\tau {\tilde n P}_+\displaystyle,
\nonumber
\end{eqnarray}
with $P_\pm = P_2 \pm P_1$.
This function depends on the distribution parameter
$t$ and the quotient $\tau = {\tilde n P}_- / {\tilde n P}_+$ of the
projection of momenta onto a light-like direction $\tilde n$, the
renormalization point $\mu$, and the scalar products of the external
moments $P_i P_j$.
For physical states $|P_1>$ and $|P_2>$ the additional variable $\tau$ is
restricted by
\begin{eqnarray}
\label{2.7}
 |\tau| = \left |{{\tilde nP_-} \over {\tilde nP_+}} \right | =\left |
{{P_-^0- P_-^{\scriptscriptstyle \| \displaystyle}} \over {P_+^0-
P_+^{\scriptscriptstyle \| \displaystyle}}}\right| \leq 1,\quad
P^{\scriptscriptstyle \| \displaystyle} = {{\vec{\tilde n}\vec
P}\over{|\vec{\tilde n}|}} .
\end{eqnarray}

With the help of the $\alpha$-representation it is possible to investigate
the support properties of the function $q^a(t,\tau ,\mu^2)$ with respect
to the variable $t$.
It turns out that $ q^a (t,\tau ,Q^2)=0 \quad \hbox{for} \quad |t|>1$.
This distribution amplitude is of relevance for e.g., the description of
nonforward processes near the forward case, where the parton picture with
two distribution functions cannot be applied.
On the other hand, it is possible to apply such a function in the
limit $P_2 \rightarrow 0$ too. Of course, then this state is not the vacuum
state, but for the mathematical aspects of the connection of evolution
equations for forward and nonforward processes this is very useful. In the
limit of forward scattering we reach the ``forward distribution amplitude".
This amplitude is real for $ P_1 = P_2 $, and for this reason  it
simultaneously plays the role of the parton distribution function
in deep inelastic scattering
(For the virtual Compton scattering amplitude approximated by a
light-cone expansion the formation of its absorptive part
is nontrivial for the hard scattering part only.).
So we can perform two limits:
\begin{eqnarray}
\label{2.9}
\Phi^a (x=(1+t)/2, Q^2)&=&
   \lim_{\tau \to -1}q^a (t,\tau ,Q^2) \qquad
\mbox{meson wave function,} \\
q^a (z,Q^2)&=& \lim_{\tau \to 0} q^a (z,\tau ,Q^2) \qquad \mbox{  quark
distribution
function}.\nonumber
\end{eqnarray}

Next we need evolution equations valid for the
distribution amplitude (\ref{2.6}). As input we can use the
renormalization group equations (\ref{3.2})
for  the  light-ray  operators (\ref{2.2}). As special cases we should
find  the Brodsky-Lepage (BL) and the Altarelli-Parisi (AP) equations.

For  the  derivation  of  the  evolution  equation  we differentiate
the general distribution amplitude $q^a(t,\tau ,\mu^2)$ with
respect to the renormalization parameter. Thereby we take into account its
representation in terms of matrix elements of light-ray operator (\ref{2.6}).
The differentiation of this operator can be performed with the help of its
renormalization group equation (\ref{3.2}). A straightforward calculation
starts from
\begin{eqnarray}
\label{4.1}
 \lefteqn{  \mu {d\over d\mu} q^a(t,\tau ,\mu^2) = } \\
& &    \int\!{d\kappa_-(\tilde nP_+)\over 2\pi (\tilde nP_+)}\;
                         e^{i\kappa_-(\tilde nP_+)t} \mu{d\over d\mu}
                         <P_2|RTO^a(\kappa_+=0,\kappa_-;\tilde n)e^{iS_I}|P_1>
         |\scriptstyle \tilde nP_- = \tau (\tilde nP_+) . \nonumber
\end{eqnarray}
We obtain the evolution equation
\begin{eqnarray}
\label{4.4}
Q^2{d\over dQ^2}q^a(t,\tau ,Q^2)=\int_{-1}^1\!{dt'\over |2\tau|}
\left(\gamma \left({t\over \tau},{t'\over \tau}\right)-2\gamma_\psi \delta
\!\left({t\over \tau} -{t'\over \tau}\right)\right) q^a(t',\tau ,Q^2)
\end{eqnarray}
 with the evolution kernel
\begin{eqnarray}
\label{4.5}
\gamma (t,t') = \int\!dw_-\; \gamma (w_+=t'w_--t,w_-),
\end{eqnarray}
\begin{eqnarray}
\label{3.23}
w_+={{\kappa_+'-\kappa_+}\over {\kappa_-}}
 ,\quad w_-={{\kappa_-'} \over {\kappa_-}},
\end{eqnarray}
where  $\gamma  (w_+,w_-)  $  is  directly  related  to  the  original
anomalous dimension $\gamma (\underline{\kappa } ,\underline{\kappa '})$.
Using translation and scale invariance we get
\begin{eqnarray}
\label{3.24}
\gamma ({\kappa }_+, {\kappa }_-  ;{\kappa '}_+,  {\kappa '}_-)
  =\frac{1}{\kappa_-^{2}} \gamma (0,1,
  {{\kappa_+'-\kappa_+}\over {\kappa_-}},{{\kappa_-'} \over {\kappa_-}})
  =\frac{1}{\kappa_-^{2}} \gamma (w_+ ,w_-).
\end{eqnarray}
This   property is essential for reducing the two-variable renormalization
group equation (\ref{3.2}) to a one-variable evolution equation (\ref{4.4}).
{}From the renormalization group invariance of
$\int dt\; q^a(t,\tau,\mu^2)$  it follows that
\begin{eqnarray}
\label{4.6}
\int\! dt\; \gamma (t,t') = 2\gamma_\psi,
\end{eqnarray}
so that the standard ``+"-definition for the generalized function
\begin{eqnarray}
\label{4.7}
[\gamma (t,t')]_+=\gamma(t,t')-\delta(t-t')\int\!    dt'' \;
\gamma (t'',t')=\gamma (t,t')-2\gamma_\psi \delta (t-t')
\end{eqnarray}
arises in a very natural way. We see that instead of the original anomalous
dimension there appears an evolution kernel that contains the original
anomalous dimension as an essential input. It
seems to be natural to denote the general evolution kernel as
an extended BL-kernel. The reason for this is the following:  restricting
$\gamma(t,t')$ to the parameter region $|t|,|t'| \leq 1$, it coincides
with  the evolution kernel for the hadron wave function, but eqs. (\ref{4.5})
and (\ref{3.24}) provide us with a meaningful definition outside this region.

In this way, we obtain  evolution equations for all forward and
nonforward matrix elements. To our knowledge,  in the literature
up to now only two examples have been investigated, the case of forward
scattering and the case of the meson wave function. Here, both cases are
contained as limits $P_2 \to P_1$ or $P_2 \to 0$. We write
these limits down in a short form:
\begin{itemize}
\item Evolution
equation for the quark distribution function (forward scattering)
with the AP-kernel $P(z/{z'})$
\begin{eqnarray} q^a (z,Q^2)&=& \lim_{\tau \to 0} q^a (z,\tau ,Q^2),
\nonumber \\
\label{4.8}
Q^2{d\over dQ^2}q^a(z,Q^2)&=&\int_{-1}^1\!{dz'\over |z'|}\;
P\!\left({z\over z'};\alpha_s(Q^2)\right) q^a(z',Q^2)\\ 
\label{4.9}
|z'|^{-1}P\!\left({z\over z'}\right)&=&\lim_{\tau \to 0} {1\over |2\tau |}
\left[\gamma\!\left({z\over \tau},{z'\over \tau}\right)\right]_+
\end{eqnarray}
\item Evolution equation for meson wave functions with the BL-kernel
$V_{BL}(x,y)$
\begin{eqnarray}
\Phi^a (x=(1+t)/2,Q^2)&=& \lim_{\tau \to -1}q^a (t,\tau ,Q^2), \nonumber\\
\label{4.10}
Q^2{d\over dQ^2} \Phi^a(x,Q^2)&=&\int_{0}^1\!dy\;
V_{BL}(x,y;\alpha_s(Q^2)) \Phi^a(y,Q^2).\\
\label{4.11}
V_{BL}(x,y)&=&[\gamma (2x-1,2y-1)]_+|\scriptstyle 0\leq x,y\leq 1,
\end{eqnarray}
where $\alpha_s(Q^2)$ denotes the running coupling constant.
Here, the variables $x=(1+t)/2$ and $y = (1+t')/2$ are restricted to
$0\leq x,y\leq 1$.
Note, however, that in fact the quantum numbers must change discontinuously,
so that the limit is only formal.
\end{itemize}

\noindent
Note that both kernels (\ref{4.9}) and (\ref{4.11}) are calculated
separately in QCD\@. From the above considerations it is obvious, however,
that they  have a common origin, namely the anomalous dimension
$\gamma(w_+,w_-)$.

\section{ The Extended BL-Kernel, Relations between the BL- and AP-Kernels}
\setcounter{equation}{0}
\renewcommand{\theequation}{\arabic{section}.\arabic{equation}}

We have obtained a new evolution kernel (extended BL-kernel) for distribution
amplitudes containing the (restricted) BL-kernel and the AP-kernel as special
cases.
The standard perturbative calculations of the BL-kernel define it as an
evolution kernel for a restricted range of the variables.
It turns out that this kernel already contains the essential information
and that an explicit continuation procedure can be prescribed.

The first question is: What is the region in the $(t,t')$-plane
where the kernel $\gamma (t,t')$ is defined?
After some algebra we obtain the following representation in the
$(t,t')$-plane \cite{DG}:
\begin{eqnarray}
\label{6.1}
  \gamma(t,t')&=&
      [\theta (t-t')\theta (1-t)-\theta (t'-t)\theta (t-1)] f(t,t') \\
   &&+[\theta (t'-t)\theta (1+t)-\theta (t-t')\theta (-t-1)] f(-t,-t')
  \nonumber\\
   &&+[\theta (-t-t')\theta (1+t)-\theta (t'+t)\theta (-t-1)] g(-t,t')
  \nonumber\\
   &&+[\theta (t+t')\theta (1-t)-\theta (-t'-t)\theta (t-1)] g(t,-t'),
  \nonumber
\end{eqnarray}
where the functions $f(t,t')$ and $g(t,t')$ are given by
\begin{eqnarray}
\label{6.2}
 f(t,t')&=&\int_0^{1-t\over 1-t'}dw_-\;\gamma (w_+=t-w_- t',w_-)
   , \nonumber \\
 g(-t,t')&=&\int_0^{1+t\over 1-t'}dw_-\;\gamma (w_+= -t-w_- t',-w_-).
\end{eqnarray}
Note, that for $|t|,|t'|>1$, because of support restrictions, the lower
boundary in the integral representations (\ref{6.2}) is not attained.
However, in these regions only the differences
   $  f(t,t')-f(-t,-t')\quad \hbox{and} \quad g(-t,t')-g(t,-t') $
appear, so that such undetermined contributions eliminate each other.
It is clear, therefore,  that the  kernel is defined in the complete
$(t,t')$-plane as it can be seen in Fig.\ 1\@.

\begin{figure}[t]
\unitlength1cm
\begin{picture}(10,11)(-1.5,0)

\thicklines

\put(8.5,5){\vector(1,0){0.8}}
\put(10,4.8){$t$}
\put(1.9,4.6){$-1$}
\put(7.7,4.6){$1$}
\put(5,9.2){\vector(0,1){0.8}}
\put(5.1,9.8){$t'$}
\put(5.2,2.0){$-1$}
\put(5.2,7.9){$1$}

\put(2.5,2.5){\line(0,1){5}}
\put(7.5,2.5){\line(0,1){5}}
\multiput(2.5,0)(0,0.2){12}{\circle*{0.09}}
\multiput(7.5,0)(0,0.2){12}{\circle*{0.09}}
\multiput(2.5,7.5)(0,0.2){13}{\circle*{0.09}}
\multiput(7.5,7.5)(0,0.2){13}{\circle*{0.09}}


\put(0,0){\line(1,1){2.5}}
\put(7.5,7.5){\line(1,1){2.5}}
\multiput(2.5,2.5)(0.2,0.2){25}{\circle*{0.09}}
\put(10,0){\line(-1,1){2.5}}
\put(2.5,7.5){\line(-1,1){2.5}}
\multiput(2.5,7.5)(0.2,-0.2){25}{\circle*{0.09}}

\put(1.2,9){$g_{+-}-$}    \put(1.7,8.5){$g_{-+}$}
\put(7.7,9){$f_{--}-$}    \put(7.7,8.5){$f_{++}$}
\put(1.7,1.4){$f_{++}$}    \put(1.3,0.9){$-f_{--}$}
\put(7.7,1.4){$g_{-+}$}     \put(7.7,0.9){$-g_{+-}$}

\put(4.1,8.9){$f_{--}+g_{+-}$}
\put(2.8,4.9){$f_{--}+g_{-+}$}
\put(5.4,4.9){$f_{++}+g_{+-}$}
\put(4.1,1.3){$f_{++}+g_{-+}$}

\put(1,6.5){$0$}
\put(9,6.5){$0$}
\put(1,3.5){$0$}
\put(9,3.5){$0$}

\thinlines
\multiput(2.5,7.5)(0.4,0){12}{\circle*{0.02}}
\multiput(2.5,2.5)(0.4,0){12}{\circle*{0.02}}

\put(7.5,2.5){\circle*{0.09}}
\put(5,2.5){\circle*{0.09}}
\put(5,7.5){\circle*{0.09}}
\put(2.5,5){\circle*{0.09}}
\put(7.5,5){\circle*{0.09}}

\end{picture}
\caption[FIG.1]{Support  of  $\gamma  (t,t')$, where
$f_{\pm\pm}=f(\pm t,\pm t')$,  and  $g_{\pm\mp}=g(\pm t,\mp t')$
are defined by eq. (\ref{6.1}).}
\end{figure}
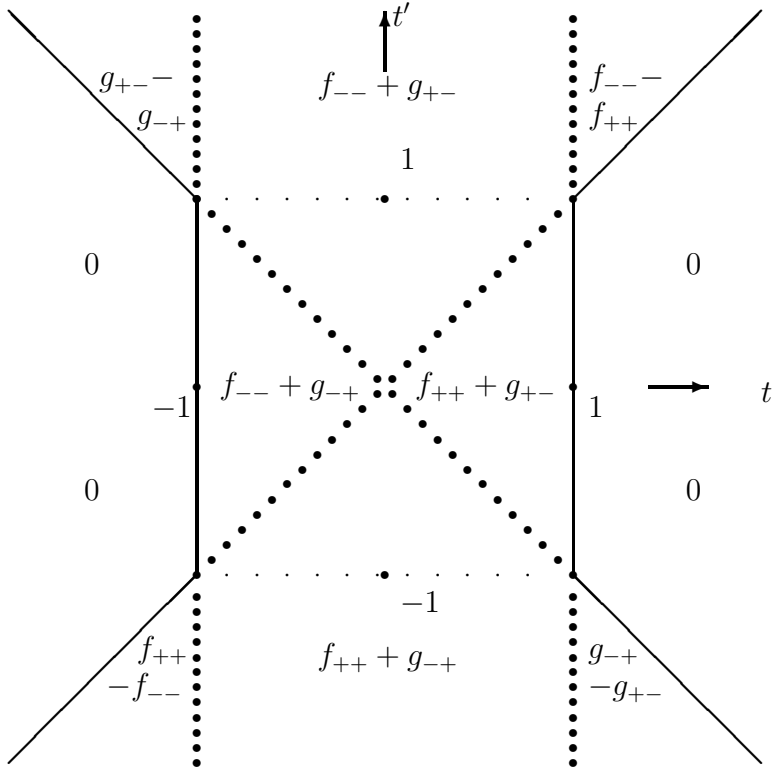
In one-loop approximation the extended BL-kernel reads
\begin{eqnarray}
\label{6.16}
   V_{BL}^{ext}(x,y)&=&\gamma_0(t=2x-1,t'=2y-1)
      = \frac{\alpha_s}{2\pi} c_F [V_0(x,y]_+    \\
   V_0(x,y)   &=&\theta\left(1-{x\over y}\right)
                               \theta\left({x\over y}\right)\hbox{sign}(y)
                               {x\over y}\left(1-{1\over x-y}\right) +
         \left\{{x\to \overline{x}\atop y\to \overline{y}}\right\}.
\nonumber
\end{eqnarray}
($c_F$ is the known group theoretic constant).
This  result  can be obtained by an extension procedure or by direct
calculation of $\gamma_0(t,t')$.

As an interesting example we shall determine the AP-kernel from the known
extended BL-kernel.
According to eq. (\ref{4.9}) it is just this new region
($|t|,|t'|>1$)  which  is  essential  for the determination of the
AP-kernel.
We first discuss the $\theta$-structure and then consider the limiting
process for the coefficient functions contained in eq. (\ref{6.16}).
The limit $\tau\to 0$, with $x=z/\tau $ and $y=1/\tau$,
for the $\theta$-functions are
\begin{eqnarray}
\label{6.25}
\left. {\theta ({x\over y})\,\theta (1-{x\over y})\atop
\theta ({1-x\over 1-y})\,\theta (1-{1-x\over 1-y})}\right\}
\quad &\to& \theta (z)\,\theta (1-z), \\
\label{6.26}
\left. {\theta ({1-x\over y})\,\theta (1-{1-x\over y})\atop
\theta ({x\over 1-y})\,\theta (1-{x\over 1-y})}\right\}
\quad &\to& \theta (-z)\,\theta (1+z).
\end{eqnarray}
Related  $\theta$-structures (structures that turn into each other under
$x\leftrightarrow \overline{x}=1-x$ and $y\leftrightarrow\overline{y}=1-y$)
have the same limit.
Putting it all together we obtain an expression for the AP-kernel which follows
from the extended BL-kernel (\ref{4.9}), (\ref{6.16}), and (\ref{6.25}):
\begin{eqnarray}
\label{6.42}
    P(z)={\alpha_s\over 2\pi}c_F[P_0(z)]_+
\end{eqnarray}
and
\begin{eqnarray}
\label{6.43}
                P_0(z)& = &\lim_{\tau\to 0}|\tau|^{-1}
              V_0\left({z\over\tau},{1\over\tau}\right)\\
          &=&\theta (z)\,\theta (1-z) {1+z^2\over 1-z}. \nonumber
\end{eqnarray}
The same procedure can be performed for the two-loop contributions to the
BL-kernel. The calculations are much more complicated.
The result obtained for the AP-kernel \cite{DG} is equivalent to
that of \cite{2A}. It also includes in a natural way the second order
contributions stemming from the internal anti-quark lines.
This  constitutes  a  consistency  check  of  the  calculation  of the
diagonal local anomalous dimensions.
The  nondiagonal  elements  of the anomalous dimension matrix have been
controlled by exploiting conformal symmetry breaking \cite{DM}.
Note, that our regularization prescription for the extended BL-kernel,
 after performing the limiting process, turns automatically into the well
accepted regularization prescription for the AP- kernel.\\

\section{ Evolution Equations and Structure Functions for Two-Photon
 Processes}
\setcounter{equation}{0}
\renewcommand{\theequation}{\arabic{section}.\arabic{equation}}

Here we consider the virtual nonforward Compton scattering amplitude
of a spinless particle (near forward direction) at a large off-shell
behavior of the incoming photon.
The considered region is a generalized Bjorken region, where (as in the case
of deep inelastic scattering) this process is dominated by contributions from
the light-cone.
The kinematics of the process are the following:
\begin{eqnarray}
  \gamma^\ast (q_1) + \hbox{H}(p_1) = \gamma^\ast(q_2) + \hbox{H}(p_2),
\end{eqnarray}
\begin{eqnarray}
  P &=& p_1 +p_2 = (2E_p,\vec 0),\qquad   q = (1/2) (q_1 +q_2),\\
  \Delta &=& (p_2 - p_1) = (q_1 - q_2)= (0, -2\vec p),\quad
   E_p = \sqrt{(m^2 + \vec p^2)}.\nonumber
\end{eqnarray}
The  last  notations  are the values of the momenta in the Breit frame.
The generalized Bjorken region is given by
\begin{eqnarray}
\nu = Pq = 2 E_p q_0\rightarrow \infty, \qquad
Q^2 = -q^2 \rightarrow \infty,
\end{eqnarray}
with the scaling variables
\begin{eqnarray}
   \xi  &=&  {-q^2  \over Pq} , \qquad \eta = {\Delta q \over Pq} = {q_1^2
   - q_2^2 \over 2\nu} = {|\vec p|\over E_p} \cos\phi .
\end{eqnarray}
To  understand  this process better we introduce the angle between the
vectors $\vec p$  and  $\vec q$ in the Breit frame by
$\cos\phi = {\vec p \vec q / (|\vec p| |\vec q|)}$.
In terms of these variables we get
\begin{eqnarray}
   q^2_1 = -\left( \xi -{|\vec p|\over E_p} \cos\phi \right) \nu,\qquad
   q^2_2 = -\left( \xi +{|\vec p|\over E_p} \cos\phi \right)\nu.
\end{eqnarray}
In contrast to deep inelastic scattering the variable $\xi$ is not
restricted to $ 0 \le \xi \le 1$.
For example $ q^2_2 =0$ demands $ \xi= -(|\vec p|/E_p)\cos\phi$.
It may be shown that in this region the helicity amplitudes
$T(\lambda',\lambda) = \varepsilon_2^\mu(\lambda')
    T_{\mu\nu} \times \varepsilon_1^\nu(\lambda)$
are given by
$(1/2)\varepsilon_2(\lambda')\varepsilon_1(\lambda) T_\mu^\mu$
for the transverse helicities and vanish otherwise. Therefore, only
the trace of the scattering amplitude
\begin{eqnarray}
T_{\mu\nu}(P,\Delta,q) =
   i \int d^4 x\, e^{iqx}\,<P_2|T\left(J_\mu\left(\frac{x}{2}\right)
                                  J_\nu\left(\frac{-x}{2}\right)\right)|P_1>
\end{eqnarray}
has to be considered;
($J_\mu(x) = (1/2):\overline{\psi}(x) \gamma_\mu
            (\lambda^3 - \lambda^8/\sqrt 3)\psi(x) : $
is the electromagnetic current of the hadrons (for flavour $SU(3)$).

In  our special case the light-ray operator product expansion contains
in leading order  only the following quark operator
\begin{eqnarray}
O^a(\tilde{x},\kappa_1,\kappa_2) =
 :\overline{\psi}(\kappa_1\tilde{x})(\tilde{x}\gamma)
 U(\kappa_1\tilde{x},\kappa_2\tilde{x})\psi(\kappa_2\tilde{x}):.
\end{eqnarray}
The singular coefficient functions $F_a$ are determined
perturbatively;  with   the  Born  approximation  of  the  coefficient
functions we obtain finally
\begin{eqnarray}
\label{T}
T^\mu_\mu(P,\Delta,q) \approx 2 \int_{-1}^{1} dt
\left(\frac{1}{\xi +t} -\frac{1}{\xi - t}\right)
e_aq^a_C(t,\eta;\mu^2 =Q^2),
\end{eqnarray}
where
$e_a =(2/9)\delta_{a0} + (1/6)\delta_{a3} + (1/6\sqrt 3)\delta_{a8}$,
and the distribution amplitudes
\begin{eqnarray}
q^a_C(t,\tau;\mu^2) =
\int \frac{d(\kappa_-\tilde{x}P)}{2\pi\tilde{x}P}
\,e^{i\kappa_-\tilde{x}P\,t}
<P_2| O^a(\tilde{x},\kappa_-,\kappa_+=0)_{\mu^2}|P_1>_{\big|\tilde{x}\Delta
=\tau\tilde{x}P}
\end{eqnarray}
contain the long range behaviour of the process (which is related to
the hadron states $|P_i>$).
This  amplitude  generalizes  the parton distribution function of deep
inelastic scattering. It has the following properties:
\begin{itemize}
\item the support of  $q^a_C(t,\tau,Q^2)$ is restricted by $|t|\leq 1$ and
$|\tau|\leq 1$,
\item $q^a(t,Q^2) = q^a_C(t,\tau =0,Q^2)$ is essentially the parton
distribution function used in deep inelastic scattering,
\item $q^a_C(t,\tau =0,Q^2)$ is real.
\end{itemize}
This  generalized  distribution  function  satisfies our new evolution
equation Eq.\ (\ref{4.4}). This conference contribution collects results of
the article \cite{HOROR}.


\end{document}